\documentclass{svmult}

\usepackage[dvips]{graphicx}
\usepackage{amssymb, amsmath}
\usepackage{makeidx}  
\usepackage{multicol} 
\usepackage[bottom]{footmisc}
\usepackage{color}

\makeindex   

\def\R{\mathbb{R}}

\begin{document}

\title*{Nonlinear Parabolic Equations arising in Mathematical Finance}
\titlerunning{Nonlinear Parabolic Equations arising in Mathematical Finance}
\author{Daniel \v{S}ev\v{c}ovi\v{c}\inst{1}}
\institute{Dept.\ Applied Mathematics \& Statistics, Comenius University, 842 48  Bratislava, Slovakia. \texttt{sevcovic@fmph.uniba.sk}}

\maketitle

\begin{abstract}

This survey paper is focused on qualitative and numerical analyses of fully nonlinear partial differential equations of parabolic type arising in financial mathematics. The main purpose is to review various non-linear extensions of the classical Black-Scholes theory for pricing financial instruments, as well as models of stochastic dynamic portfolio optimization leading to the Hamilton-Jacobi-Bellman (HJB) equation. After suitable transformations, both problems can be represented by solutions to nonlinear parabolic equations.  Qualitative analysis will be focused on issues concerning the existence and uniqueness of solutions. In the numerical part we discuss a stable finite-volume and finite difference schemes for solving fully nonlinear parabolic equations.

\end{abstract}


\noindent\textit{Keywords and phrases}{
Option pricing, nonlinear Black-Scholes equation}

\section{Nonlinear generalization of the Black-Scholes equation for pricing financial instruments}

According to the classical theory developed by  Black,  Scholes and Merton the value $V(S,t)$ of an option in the idealized financial market can be computed from a solution to the well-known Black--Scholes linear parabolic equation:
\begin{equation}
\label{eq6}
\partial _t V + \frac{1}{2}\sigma^2 S^2\partial _S^2 V + (r - 
q)S\partial _S V - rV = 0,\mbox{ }t \in [0,T),S > 0,
\end{equation}
derived by Black and Scholes and, independently by Merton (c.f. \cite{NBS5},\cite{NBS4}). Here $\sigma>0$ is the volatility of the underlying asset driven by the geometric Brownian motion, $r > 0$ is the risk-free interest rate of zero-coupon bond and $q \ge 0$ is the dividend rate. Similarly, as in the case of the HJB equation the solution is subject to the terminal condition $V(S,T) = \bar {V}(S)$ at $t = T$.

The linear Black--Scholes equation with constant volatility $\sigma$ has been derived under several restrictive assumptions like  e.g., frictionless, liquid and complete markets, etc. We also recall that the linear Black--Scholes equation provides a solution corresponding to a perfectly replicated hedging portfolio which need not be a desirable property. In the last decades some of these assumptions have been relaxed in order to model, for instance, the presence of transaction costs (see e.g. Leland \cite{NBS5,NBS12} and Avellaneda and Paras \cite{NBS7}), feedback and illiquid market effects due to large traders choosing given stock-trading strategies (Sch\"onbucher and Willmott \cite{NBS13}, Frey and Patie \cite{NBS11}, Frey and Stremme \cite{NBS10}, imperfect replication and investor's preferences (Barles and Soner \cite{NBS8}), risk from the unprotected portfolio (Janda\v{c}ka and \v{S}ev\v{c}ovi\v{c} \cite{NBS1}). Another nonlinear model in which 
transaction costs are described by a decreasing function of the number of shares has been derived by Amster \textit{et al.} \cite{NBS6}. In all aforementioned generalizations of the linear BS equation (\ref{eq6}) the constant volatility $\sigma$ is replaced by a nonlinear function:

\begin{equation}
\label{eq7}
\sigma= \sigma(S \partial _S^2 V)
\end{equation}
depending on the second derivative $\partial _S^2 V$ of the option price itself. 

One of the first nonlinear models taking into account transaction costs is the Leland model for pricing the call and put options. This model was further extended by Hoggard, Whalley and Wilmott \cite{NBS12} for general type of derivatives. In this model the variance $\sigma^2$ is given by
\begin{equation}\label{sigma_Leland}
\sigma(S\partial^2_S V)^2 = \sigma^2_0
\left(1-\mathrm{Le \, sgn}\left(S\partial_S^2 V \right)\right)=\left\{ 
\begin{array}{r@ {\quad}l}
    \sigma^2(1-\mathrm{Le}), & \mathrm{if} \, \partial_S^2 V>0, \\
    \sigma^2(1+\mathrm{Le}), & \mathrm{if} \, \partial_S^2 V<0, \\   
\end{array} \right.
\end{equation}
where $\mathrm{Le} =   \sqrt{\frac{2}{\pi}} \frac{C_0}{\sigma \sqrt{\Delta t}}$ is the so-called Leland number, $\sigma_0$ is a constant historical volatility, $C_0>0$ is a constant transaction costs per unit dollar of transaction in the underlying asset market and $\Delta t$ is the time--lag between consecutive portfolio adjustments. The nonlinear model with the volatility function given as in  (\ref{sigma_Leland}) can be also viewed as a jumping volatility model investigated by Avellaneda and Paras \cite{NBS7}. 

The important contribution in this direction has been presented in the paper \cite{NBS6} by Amster, Averbuj, Mariani and Rial, where the transaction costs are assumed to be a non-increasing linear function of the form $C(\xi)=C_0-\kappa\xi$, ($C_0, \,\kappa >0$), depending on the volume of trading stock $\xi\geq0$ needed to hedge the replicating portfolio. A disadvantage of such a transaction costs function is the fact that it may attain negative values when the amount of transactions exceeds the critical value $\xi = C_0/\kappa$. In the model studied by Amster \emph{et al.} \cite{NBS6}  (see also Averbuj \cite{Averbuj2012}, Mariani \emph{et al.} \cite{Mariani2011}) volatility function has the following form:

\begin{equation}\label{sigma:Amster}
\sigma(S\partial^2_S V)^2=\sigma^2_0 \left( 1 -  \mathrm{Le}\, \mathrm{sgn} \left(S\partial_S^2 V \right) + \kappa  S \partial_S^2 V \right).
\end{equation}

In the recent paper \cite{NBS14} \v{S}ev\v{c}ovi\v{c} and \v{Z}itnansk\'a investigated a model for pricing option under variable transaction costs. 
\begin{equation}\label{sigma:variable}
\sigma(S\partial^2_S V)^2=\sigma^2_0
\left( 1  - \sqrt{\frac{2}{\pi}} \tilde C(\sigma S |\partial_S^2 V| \sqrt{\Delta t}) 
\frac{\mathrm{sgn}( S \partial_S^2 V)}{\sigma\sqrt{\Delta t}}
\right)
\end{equation}
where $\tilde C$ is the mean value modification of the transaction cost function $C=C(\xi)$ defined as follows: $\tilde C(\xi) = \int_0^\infty C(\xi x) x\, e^{-x^2/2}  dx$. As an example one can consider the piecewise linear transaction cost function of the form:
\begin{equation}\label{nonlin:TCPiecewise}
C(\xi) = \left\{ \begin{array}{l@ {\quad} l @{\quad} r}
     C_0,                         &  \mbox{if } &\,     0\le \xi \le \xi_-, \\
     C_0 - \kappa (\xi - \xi_-),  &  \mbox{if } &\, \xi_-\le \xi \le \xi_+, \\
     \underline{C}_0,             &  \mbox{if } &\,          \xi \ge \xi_+. \\
\end{array} \right.
\end{equation}

In \cite{BaHowison} Bakstein and Howison investigated a parametrized model for liquidity effects arising from the asset trading. In their model $\sigma$ is a quadratic function of the term $H=S\partial_S^2 V$:
\begin{align}
\sigma(S\partial^2_S V)^2= & \sigma^2_0 \Biggl(  1+\bar{\gamma}^2(1-\alpha)^2 + 2\lambda  S \partial^2_S V +  \lambda^2(1-\alpha)^2\left(S\partial^2_S V\right)^2  
\nonumber \\ 
&+ 2\sqrt{\frac{2}{\pi}} \bar{\gamma} \, \mathrm{sgn}\left(S\partial^2_S V\right) + 2\sqrt{\frac{2}{\pi}} \lambda (1-\alpha)^2 \bar{\gamma}  \left|S \partial^2_S V \right| \Biggr).
\end{align}
The parameter $\lambda$ corresponds to a market depth measure, i.e. it scales the slope of the average transaction price. Next, the parameter $\bar{\gamma} $ models the relative bid--ask spreads and it is related to the Leland number through relation $2\bar{\gamma}\sqrt{2/\pi}=\mathrm{Le}$. Finally, $\alpha$ transforms the average transaction price into the next quoted price, $0\leq \alpha \leq 1 $.

The risk adjusted pricing methodology (RAPM) model takes into account risk from the unprotected portfolio was proposed by Kratka \cite{Kr}. It was generalized and analyzed by Janda\v{c}ka and \v{S}ev\v{c}ovi\v{c} in \cite{NBS1}. In this model the volatility function has the form:
\begin{equation} \label{vol_RAPM}
\sigma(S\partial^2_S V)^2=\sigma^2_0
\left(1+\mu\left(S\partial_S^2 V \right)^{\frac{1}{3}} \right),
\end{equation}
where $\sigma_0>0$ is the constant historical volatility of the asset price return and $\mu=3(C_0^2R/2\pi)^{\frac{1}{3}}$, where $C_0,\,R\geq 0$ are non--negative constants representing the transaction cost measure and the risk premium measure, respectively. 

If transaction costs are taken into account perfect replication of the contingent claim is no longer possible and further restrictions are needed 
in the model. By assuming that investor's preferences are characterized by an exponential utility function   Barles and Soner (c.f. \cite{NBS8}) derived a nonlinear Black--Scholes equation with the volatility $\sigma$ given by
\begin{equation}
\sigma(S\partial^2_S V, S, t)^2 
= \sigma^2_0 \left(1+\Psi(a^2 e^{r (T-t)} S^2\partial^2_S V)\right)
\label{c-barles}
\end{equation}
where $\Psi$ is a solution to the ODE: 
\[
\Psi^\prime(x) = (\Psi(x)+1)/(2 \sqrt{x\Psi(x)} -x), \Psi(0)=0,
\]
and $a>0$ is a given constant representing risk aversion. Notice that $\Psi(x)=O(x^\frac{1}{3})$ for $x\to 0$ and $\Psi(x)=O(x)$ for $x\to\infty$. 

All the nonlinear volatility models mentioned in this section can be written in the form of a solution to the fully nonlinear parabolic equation:
\begin{equation}
\label{nonlin}
\partial _t V + \frac{1}{2}\sigma(\partial^2_S V)^2 S^2\partial_S^2 V + (r - 
q)S\partial _S V - rV = 0,\mbox{ }t \in [0,T),S > 0.
\end{equation}

In \cite{NBS1} Janda\v{c}ka and \v{S}ev\v{c}ovi\v{c} proposed the method of transformation of equation (\ref{nonlin}) into a quasi-linear parabolic equation for the second derivative $\partial _S^2 V$ (the so-called Gamma of an option) of a solution. Indeed, if we introduce the new variables $H(x,\tau ) = S\partial _S^2 V(S,t), x = \ln S$ and $\tau = T - t$ then equation (\ref{nonlin}) can be transformed into the so-called Gamma equation:
\begin{equation}
\label{eq10}
\partial _\tau H = \partial _x^2 \beta (H) + \partial _x \beta (H) + (r - q)\partial _x H - qH,\quad x \in R,\tau \in (0,T),
\end{equation}
where
\[
\beta(x,H) = \frac12 \sigma(H)^2 H
\]
(c.f. \cite{NBS1},\cite{NBS9}). Recall that the Gamma equation can be obtained by twice differentiation with respect to $x$ of the Black--Scholes equation (\ref{eq61}) with the volatility of the general type (\ref{eq7}). A solution $H(x,\tau )$ to the Cauchy problem for (\ref{eq10}) is subject to the initial condition $H(x,0) = H_0 (x)$. 

\section{Nonlinear Hamilton-Jacobi-Bellman equation and optimal allocation problems}

Optimal allocation and optimal investment problems with state constraints  attracted a lot of attention from both theoretical as well as application point of view. The main purpose is to maximize the total expected discounted  utility of consumption for the optimal portfolio investment consisting of several stochastic assets, over infinite or finite time horizons. It is  known that the value function of the underlying stochastic control problem 
is the unique smooth solution to the corresponding Hamilton-Jacobi-Bellman (HJB) equation and the optimal consumption and portfolio are presented in feedback form (Zariphopoulou \cite{HJB10}). 

Let us consider the stylized financial market in which the aim of a portfolio manager is to maximize the expected value of the terminal wealth of a portfolio, measured by a prescribed utility function $U$. In particular, if $n$ is the number of assets entering the portfolio, $T$ the investment horizon, the goal is to  find an optimal trading strategy $\{\theta \} = \{\theta _t \in \R^n\,\vert \,t \in [0,T]\}$ belonging to a set $A = A_{0,T} $ of strategies  $A_{t,T} = \{\{\theta \}\vert \theta _s \in S^n,s \in [t,T]\},$ where $S^n = \{\theta _t \in \R^n\vert \theta _t \in [0,1]^n,1^T\theta _t = 1\}$ is a convex compact simplex such that $\{\theta \}$ maximizes the expected terminal utility from the portfolio:

\begin{equation}
\label{eq1}
\max_{\{\theta \} \in A} {\mathbb E}\left[ {U(X_T^\theta )\vert X_0^\theta = x_0 } \right].
\end{equation}
Here $X_t = \mbox{ln }Y_t $ represents a stochastic process governed by the following stochastic differential equation
\[
\mbox{d}X_t^\theta = \left( {\mu (\theta ) - \frac{1}{2}\sigma (\theta )^2} \right)\mbox{d}t + \sigma (\theta )\mbox{d}W_t 
\]
for a logarithmic portfolio value, where $x_0 $ is its initial value at the time $t = 0$. Here $\mu (\theta )$ and $\sigma (\theta )$ are the expected return and volatility of the portfolio. As a typical example, one can consider functions $\mu (\theta ) = \mu^T \theta $ and $\sigma^2(\theta ) = \theta^T\Sigma \theta $, where $\mu $ is a vector of mean returns and $\Sigma $ is a covariance matrix. It is known from the theory of stochastic dynamic programming that the so-called value function 
\begin{equation}
\label{eq2}
V(x,t): = \mathop {\sup }\limits_{\{\theta \} \in A_{t,T} } {\mathbb E}\left[ {U(X_T^\theta )\vert X_t^\theta = x} \right]
\end{equation}
subject to the terminal condition $V(x,T): = U(x)$ can be used for solving the stochastic dynamic optimization problem (\ref{eq1}) (c.f. Bertsekas \cite{HJB15}, Fleming and Soner \cite{HJB7}). Moreover, it is also known, that the value function $V = V(x,t)$ satisfies the following Hamilton-Jacobi-Bellman equation: 

\begin{equation}
\label{eq3}
\partial_t V + \max_{\theta \in S^n} \left\{ {\left( {\mu (\theta ) - \frac{1}{2}\sigma (\theta )^2} \right)\partial _x V + \frac{1}{2}\sigma (\theta )^2\partial _x^2 V} \right\} = 0\,,
\end{equation}
for all $x \in \R,t \in [0,T)$ and it satisfies the terminal condition $V(.,T): = U(.)$ (see e.g. \cite{HJB1,HJB3}). 

In general, explicit solutions to HJB equations are not available and this is why various numerical approaches have to be adopted. Regarding numerical approaches for solving HJB equations associated with portfolio optimization, we can mention and refer to finite difference methods for approximating its viscosity solution developed and analyzed by Tourin and Zariphopoulou \cite{HJB8}, Crandall, Ishii and Lions \cite{HJB6}, Nayak and Papanicolaou \cite{HJB12}. Other approach based on Markov chain approximation techniques was investigated by Song \cite{HJB13} and Fleming and Soner \cite{HJB7}. Classical methods for solving HJB equations are discussed by Benton in \cite{HJB10}. In \cite{HJB14}, Musiela and Zariphopoulou applied the power-like transformation in order to linearize the non-linear PDE for the value function in the case of an exponential utility function. Muthamaran and Sunil \cite{HJB16} solved a multi-dimensional portfolio optimization problem with transaction costs. They used finite element method and iterative procedure that converts a free-boundary problem into a sequence of fixed boundary problems. In \cite{NM7}, Peyrl \textit{et al.} applied a successive approximation algorithm for solving the corresponding HJB equation. The fixed point-policy iteration scheme for solving discretized HJB equations is discussed in Huang \textit{et al.} \cite{NM20}. In \cite{NM19}, Witte and Reisinger presented a penalty approach for the numerical solution of discrete continuously controlled HJB equations. 

In the recent paper \cite{HJB9} Kilianov\'a and \v{S}ev\v{c}ovi\v{c} transformed the fully nonlinear HJB equation (\ref{eq3}) into the Cauchy problem for the quasi-linear parabolic equation: 
\begin{eqnarray}
\label{eq4}
 & \partial _t \varphi + \partial _x^2 \beta (\varphi ) + \partial _x [(1 - 
\varphi )\beta (\varphi )] = 0,\quad x \in R,t \in [0,T),
\\
 & \varphi (x,T) = 1 - U^{\prime \prime }(x) / U^\prime (x),\quad x \in R.
\end{eqnarray}
To this aim we introduced the following transformation: 
\[
\varphi (x,t) = 1 - \frac{\partial _x^2 V(x,t)}{\partial _x V(x,t)}.
\]
It is referred to as the Riccati transformation and it has been proposed and studied in \cite{HJB4,HJB3} and further analyzed by Ishimura and \v{S}ev\v{c}ovi\v{c} in \cite{HJB1}. The resulting equation was solved numerically by an iterative method based on the finite volume approximation. Furthermore, it follows from the analysis \cite{HJB9} by Kilianov\'a and \v{S}ev\v{c}ovi\v{c} that the diffusion function $\beta(\varphi )$ is the value function of the following parametric optimization problem: 

\begin{equation}
\label{eq5}
\beta(\varphi ) = \min_{\theta \in S^n} \{ - \mu (\theta ) + \frac{\varphi }{2}\sigma (\theta )^2\}\,.
\end{equation}

The dispersion function $\theta \mapsto \sigma (\theta )^2$ is assumed to be strictly convex and $\theta \mapsto \mu (\theta )$ is a linear function. Therefore problem (\ref{eq5}) belongs to a class of parametric convex optimization problems (c.f. Bank \textit{et al.} \cite{NM9}, Hamala and Trnovsk\'a \cite{NM10}). Useful generalization of the HJB equation (\ref{eq3}) in case the covariance matrix $\Sigma$ belongs to some set $P$ of (e.g. ellipsoidal sets) of covariance matrices was studied Kilianov\'a and Trnovsk\'a in \cite{HJB17} with regard to application to the so-called ,,worst case variance'' portfolio model in which the diffusion function (\ref{eq5}) has the form:
\[
\beta(\varphi ) = \min_{\theta \in S^n} \max_{\Sigma \in P} { - \mu ^T\theta + \frac{\varphi }{2}\theta ^T\Sigma \theta }\,.
\]
They showed this problem can be analyzed by the methods of semidefinite programming. The value function $\beta(\varphi )$ need not be sufficiently smooth and its second derivative can have jumps. 

In fact, the Riccati transformation is the logarithmic derivative of the  derivative of the value function. In the context of a class of HJB equations 
with range constraints, the Riccati transformation has been analyzed  recently by Ishimura and \v{S}ev\v{c}ovi\v{c} in \cite{HJB1} where a traveling wave  solution to the HJB equation was constructed. Concerning numerical methods for solving the quasi-linear parabolic PDE obtained from the HJB  equation by means of the Riccati transformation we mention recent papers by  Ishimura, Koleva and Vulkov \cite{NM17, NM18}. In \cite{NM17}, Koleva considered a similar nonlinear parabolic equation, obtained by means of a Riccati-like  transformation of the Hamilton-Jacobi-Bellman equation, arising in pension saving management. In contrary to our model problem, she considered a  problem without constraints on the optimal decision. She applied two iterative numerical methods, namely the fully implicit Picard method and the mixed Picard-Newton method and discussed their accuracy and effectiveness.

In summary, the nonlinear volatility generalization of the Black-Scholes equation as well as the Hamilton-Jacobi-Bellman equation can be transformed into the quasilinear parabolic equation for the unknown function $H=H(x,\tau)$ representing either the Gamma of the portfolio  $H=S\partial^2_S V$ (nonlinear volatility Black-Scholes models) or the relative risk aversion function $H = 1 - \frac{\partial_x^2 V}{\partial_x V}$ (Hamilton-Jacobi-Bellman equation). The resulting quasilinear parabolic equation has the form:
\begin{equation}
\label{eq61}
\partial _\tau H = \partial _x^2 \beta (H) + f(x, H, \partial _x H),\quad x \in R,\tau \in (0,T),
\end{equation}
where $\beta $ is a suitable nonlinear function.

\section{Existence of classical solutions, comparison principle}

In this section we recall results on existence of classical smooth solutions to the Cauchy problem for the  quasilinear parabolic equation (\ref{eq61}). Following the methodology based on the so-called Schauder's type of estimates (c.f. Ladyzhenskaya \emph{et al.} \cite{LSU}), we shall proceed with a definition of function spaces we will work with. Let $\Omega=(x_L, x_R)\subset\R$ be a bounded interval. We denote $Q_T =\Omega\times (0,T)$ the space-time cylinder. Let $0<\lambda<1$. By $\mathcal{H}^{\lambda}(\Omega)$ we denote the Banach space consisting of all continuous functions $H$ defined on $\bar\Omega$ which are $\lambda$-H\"older continuous. It means that their H\"older semi-norm $\langle H \rangle^{(\lambda)} = \sup_{x,y\in\Omega, x\not= y} |H(x) - H(y)|/|x-y|^\lambda$ is finite. The norm in the space $\mathcal{H}^{\lambda}(\Omega)$ is then the sum of the maximum norm of $H$ and the semi-norm $\langle H \rangle^{(\lambda)}$. The space $\mathcal{H}^{2+\lambda}(\Omega)$ consists of all twice continuously differentiable functions $H$ in $\bar\Omega$ whose second derivative $\partial_x^2 H$ belongs to $\mathcal{H}^{\lambda}(\Omega)$. The space $\mathcal{H}^{2+\lambda}(\R)$ consists of all functions $H:\R\to\R$ such that $H\in \mathcal{H}^{2+\lambda}(\Omega)$ for any bounded domain $\Omega\subset\R$. 

The parabolic H\"older space  $\mathcal{H}^{\lambda, \lambda/2}(Q_T)$ of functions defined on a bounded cylinder $Q_T$ consists of all continuous functions $H(x,\tau)$ in $\bar{Q}_T$  such that $H$ is $\lambda$-H\"older continuous in the $x$-variable and $\lambda/2$-H\"older continuous in the $t$-variable. The norm is defined as the sum of the maximum norm and corresponding H\"older semi-norms. The space $\mathcal{H}^{2+\lambda, 1+\lambda/2}(Q_T)$ consists of all continuous functions on $\bar{Q}_T$ such that $\partial_\tau H, \partial^2_x H \in \mathcal{H}^{\lambda, \lambda/2}(Q_T)$. Finally, the space $\mathcal{H}^{2+\lambda, 1+\lambda/2}(\R\times [0,T])$ consists of all functions $H:\R\times [0,T]\to \R$ such that $H \in \mathcal{H}^{2+\lambda, 1+\lambda/2}(Q_T)$ for any bounded cylinder $Q_T$ (c.f. \cite[Chapter I]{LSU}).

In the nonlinear models discussed in the previous sections one can derive useful lower and upper bounds of a solution $H$ to the Cauchy problem (\ref{eq61}). The idea of proving upper and lower estimates for $H(x,\tau)$ is based on construction of suitable sub- and super-solutions to the parabolic equation (\ref{eq61}) (c.f. \cite{LSU}).
\[
\lambda_- \le \beta^\prime(H) \le \lambda_+ 
\]
for any $H\ge 0$ where $\lambda_\pm>0$ are constants. This implies strong parabolicity of the governing nonlinear parabolic equation.

\begin{theorem}\cite[Theorem 3.1]{NBS14}\label{existence}
Suppose that the initial condition $H(.,0)\ge 0$ belongs to the H\"older space $\mathcal{H}^{2+\lambda}(\R)$ for some $0<\lambda< \min(1/2,\varepsilon)$ and  $\overline{H} =\sup_{x\in\R} H(x,0) <\infty$. Assume that $\beta, f\in C^{1,\varepsilon}$ and $\beta$ satisfies $\lambda_- \le \beta^\prime(H) \le \lambda_+$ for any $0\le H\le \overline{H}$ where $\lambda_\pm>0$ are constants. 

Then there exists a unique classical solution $H(x,\tau)$ to the quasilinear parabolic equation (\ref{eq61}) satisfying the initial condition  $H(x,0)$. The function $\tau\mapsto \partial_\tau H(x,\tau)$ is $\lambda/2$-H\"older continuous for all $x\in\R$ whereas $x\mapsto\partial_x H(x,\tau)$ is Lipschitz continuous for all $\tau\in[0,T]$. Moreover, $\beta(H(.,.))\in \mathcal{H}^{2+\lambda, 1+\lambda/2}(\R\times [0,T])$ and $0<H(x,\tau) \le \overline{H}$ for all $(x,\tau)\in\R\times[0,T)$.
\end{theorem}

The proof is based on the so-called Schauder's theory on existence and uniqueness of classical H\"older smooth solutions to a quasi-linear parabolic equation of the form (\ref{eq61}). It follows the same ideas as the proof of \cite[Theorem 5.3]{HJB9} where Kilianov\'a and \v{S}ev\v{c}ovi\v{c} investigated a similar quasilinear parabolic equation obtained from a nonlinear Hamilton-Jacobi-Bellman equation in which a stronger assumption $\beta\in C^{1,1}$ is assumed.

\section{Numerical full space-time discretization scheme for solving the Gamma equation}
\label{sec5:fdm}

In this section we present an efficient numerical scheme for solving the Gamma equation. The construction of numerical approximation of a solution $H$ to (\ref{eq61}) is based on a derivation of a system of difference equations corresponding to (\ref{eq61}) to be solved at every discrete time step. We make use of the numerical scheme adopted from the paper by Janda\v{c}ka and \v{S}ev\v{c}ovi\v{c} \cite{NBS1} in order to solve the Gamma  equation (\ref{eq61}) for a general function $\beta=\beta(H)$ including, in particular, the case of the model with variable transaction costs.  The efficient numerical discretization is based on the finite volume approximation of the partial derivatives entering (\ref{eq61}). The resulting scheme is semi--implicit in a finite--time difference approximation scheme. 

Other finite difference numerical approximation schemes are based on discretization of the original fully nonlinear Black--Scholes equation in non-divergence form. We refer the reader to recent publications by Ankudinova and Ehrhardt \cite{AE}, Company \emph{et al.} \cite{CompanyNavaro}, D\"uring \emph{et al.} \cite{DFJ}, Liao and Khaliq \cite{NM12}, Zhou \emph{et al.} \cite{Zhou2015}. Recently, a quasilinearization technique for solving the fully nonlinear parabolic equation was proposed and analyzed by Koleva and Vulkov \cite{Koleva}. Our approach is based on a solution to the quasilinear Gamma equation written in the divergence form, so we can use existing finite volume based numerical scheme to solve the problem efficiently (c.f. Janda\v{c}ka and \v{S}ev\v{c}ovi\v{c} \cite{NBS1}, K\'utik and Mikula \cite{NM6}).

For numerical reasons we restrict the spatial interval to $x\in(-L,L)$ where $L>0$ is sufficiently large. Since $S=E e^x \in (E e^{-L}, E e^L)$ it is sufficient to take $L\approx 2$ in order to include the important range of values of $S$. For the purpose of construction of a numerical scheme, the time interval $[0,T]$ is uniformly divided with a time step $k=T / m$ into discrete points $\tau_j= jk$, where $j=0,1, \cdots, m$. We consider the spatial interval $[-L,L]$ with uniform division with a step  $h=L/n$, into discrete points $x_i = ih,$ where$\ i=-n,\cdots,n$.

The proposed numerical scheme is semi--implicit in time. Notice that the term $\partial_x^2\beta,$  can be expressed in the form $\partial_x^2\beta = \partial_x\left( \beta^\prime(H) \partial_x H \right)$, where $\beta^\prime$ is the derivative of  $\beta(H)$ with respect to $H$. In the discretization scheme,  the nonlinear terms $\beta^\prime(H)$ are evaluated from the previous time step $\tau_{j-1}$ whereas linear terms are solved at the current time level.

Such a discretization scheme leads to a  solution of a tridiagonal system of linear equations at every discrete time level.  First, we replace the time derivative by the time difference, approximate $H$ in nodal points by the average value of neighboring segments, then we collect all linear terms at the new time level $\tau_j$ and by taking all the remaining terms from the previous time level $\tau_{j-1}$. We obtain a tridiagonal  system for the solution vector 
$H^j=(H^j_{-n+1}, \cdots, H^j_{n-1})^T \in \R^{2n -1}$:
\begin{equation} \label{doplnky-jam-tridiagonal-bs}
a_{i}^j H_{i-1}^j+b_{i}^j H_{i}^j+c_{i}^j H_{i+1}^j = d_i^j,
\quad H_{-n}^j= 0,\ \ H_n^j =0 \,,
\end{equation}
where $i=-n+1,\cdots,n-1$ and $j=1, \cdots, m$. The coefficients of the tridiagonal matrix are given~by
\[
a_i^j = -\frac{k}{h^2}\beta^\prime_H(H_{i-1}^{j-1}) + \frac{k}{2h}r\,\quad
c_i^j = -\frac{k}{h^2}\beta^\prime_H(H_{i}^{j-1}) - \frac{k}{2h}r\,, \quad 
b_i^j = 1 - (a_i^j + c_i^j)\,,
\]
\[
d_i^j = H_{i}^{j-1} + \frac{k}{h}\Big(\beta(H_{i}^{j-1}) - \beta(H_{i-1}^{j-1} )\Big)\,.
\]
It means that the  vector $H^j$ at the time level $\tau_j$ is a solution to the system of linear equations ${\bf A}^{(j)}\, H^j = d^{j},$ where the $(2n-1)\times (2n-1)$ matrix ${\bf A}^{(j)}=\mbox{tridiag}(a^j,b^j,c^j)$. In order to solve the tridiagonal system in every time step in a fast and  effective way, we can use the efficient Thomas algorithm.

In \cite{NBS14} the authors showed that the option price  $V(S,T - \tau_j)$ can be constructed from the discrete solution $H^j_i$ by means of a simple integration scheme:
\begin{eqnarray}
\hbox{(call option)}\qquad\qquad V(S,T - \tau_j) &=&  h \sum_{i=-n}^n (S- E e^{x_i} )^+ H_i^j, \quad j=1, \cdots, m,
\nonumber \\
\hbox{(put option)}\qquad\qquad V(S,T - \tau_j) &=&  h \sum_{i=-n}^n (E e^{x_i} -S )^+ H_i^j, \quad j=1, \cdots, m.
\nonumber
\end{eqnarray}

\section{Numerical results for the nonlinear model with variable transaction costs}

\begin{figure}
\begin{center}
\includegraphics[width=0.45\textwidth]{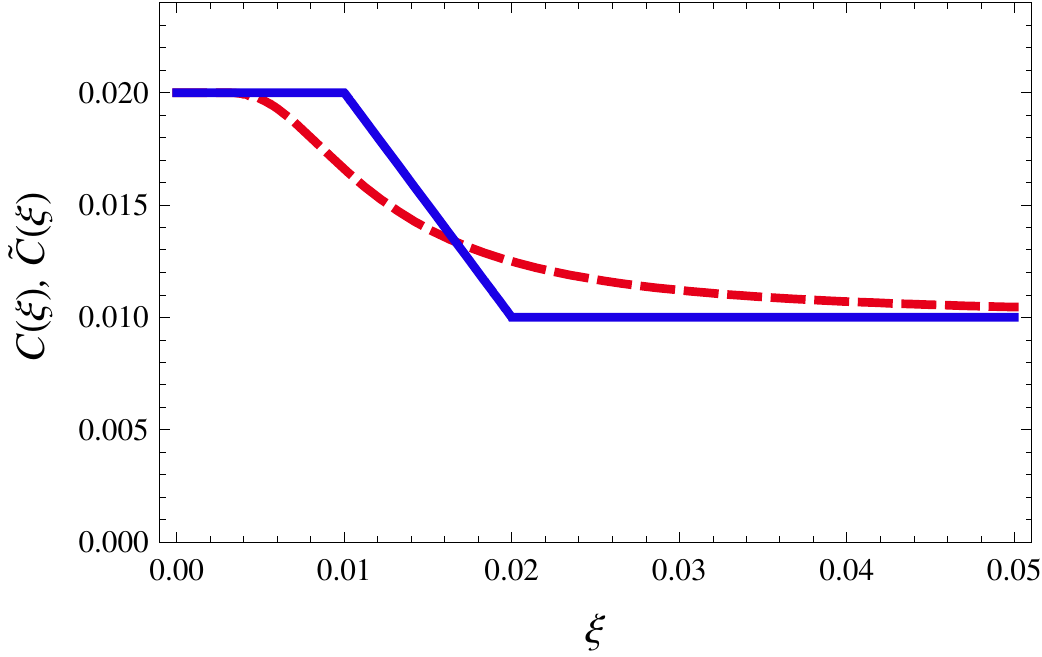}
\includegraphics[width=0.45\textwidth]{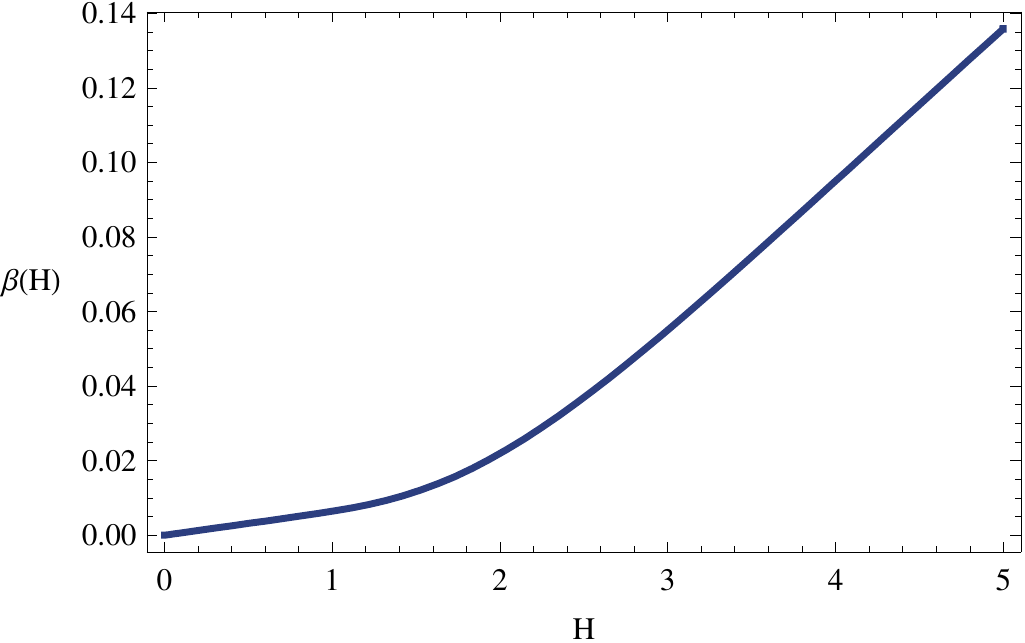}
\end{center}

\caption{%
Left: The piecewise linear transaction costs function $C$ (solid line), its mean value modification $\tilde C$ (dashed line). Right: the graph of the corresponding function $\beta(H)$. Source \cite{NBS14}
}
\label{fig:Cfun}
\end{figure}

In this section we present the numerical results for computation of the option price for the nonlinear volatility Black-Scholes model with variable transaction costs derived and analyzed by \v{S}ev\v{c}ovi\v{c} and \v{Z}itnansk\'a in the recent paper \cite{NBS14}. As an example for numerical approximation of a solution we consider variable transaction costs described by the piecewise linear non-increasing function, depicted in Figure~\ref{fig:Cfun}. The function $\beta(H)$ corresponding to the variable transaction costs function $C(\xi)$ has the form 
\begin{equation*}
\beta(H) = \frac{\sigma^2_0}{2} \left( 1 - \sqrt{\frac{2}{\pi}}\tilde{C}(\sigma|H|\sqrt{\Delta t}) \frac{\mathrm{sgn}(H)}{\sigma\sqrt{\Delta t}} \right) H,
\end{equation*}
where $\tilde C$ is the modified transaction costs function.

\begin{figure} 
\begin{center}
\includegraphics[width=0.48\textwidth]{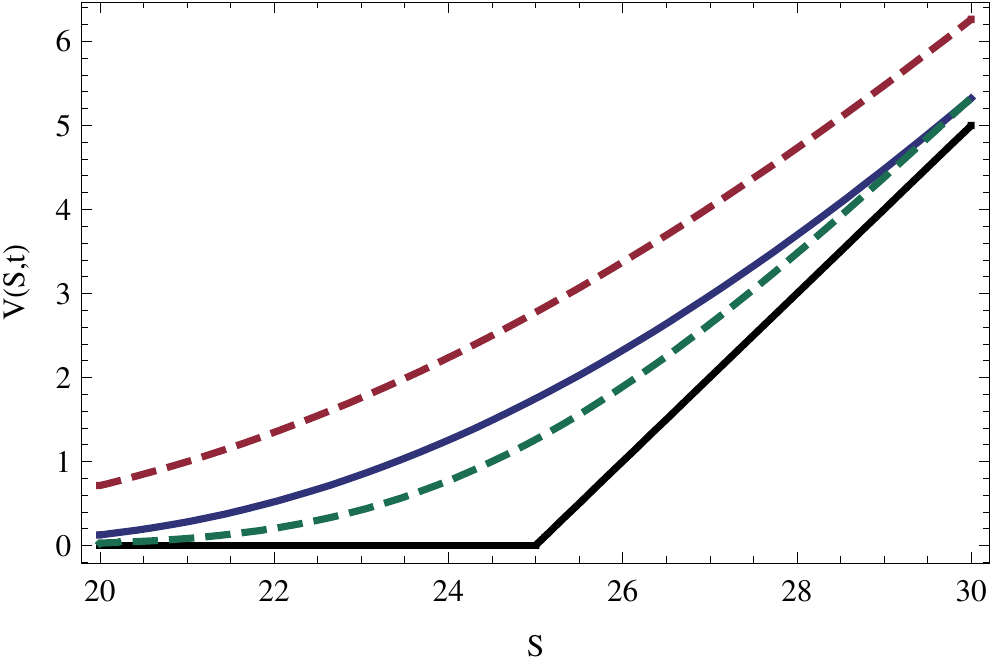}
\quad
\includegraphics[width=0.48\textwidth]{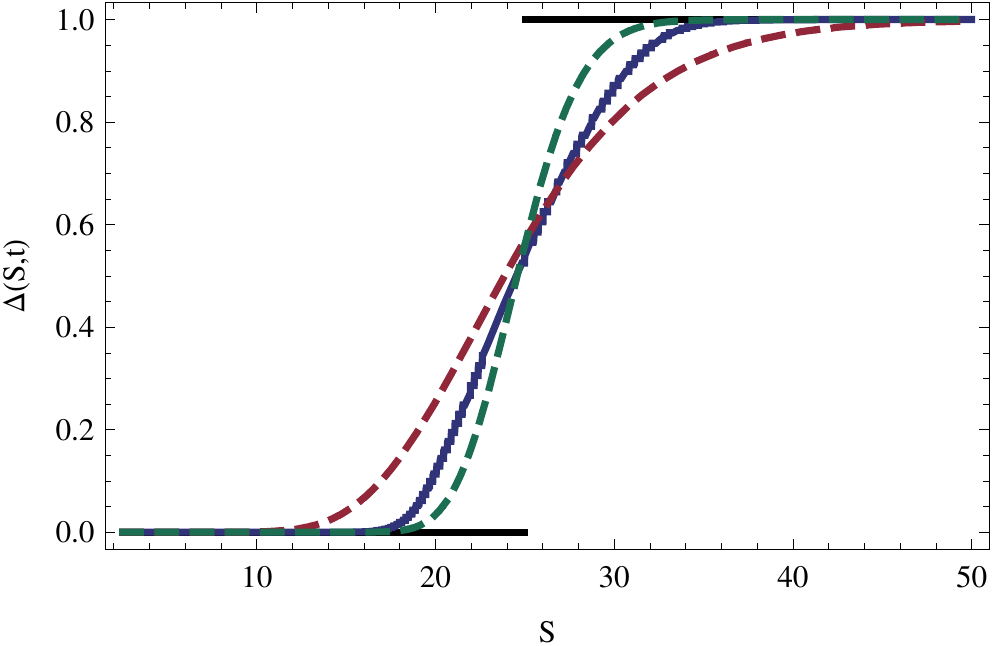}
\caption{
The call option price $V(S,t)$ as a function of $S$ for $t=0$ (left) and its delta  $\Delta(S,t)=\partial_S V(S,t)$. Source \cite{NBS14}
}
\label{fig:nonlinearTC-SolutionDelta}
\end{center}
\end{figure}

In our computations we chose the following model parameters describing the piecewise transaction costs function: $C_0 = 0.02,  \kappa = 0.3, \xi_- = 0.05, \xi_+ = 0.1$. The length of the time interval between two consecutive portfolio rearrangements: $\Delta t = 1/261$. The maturity time $T = 1$, historical volatility $\sigma = 0.3$ and the risk-free interest rate $r = 0.011$. As for the numerical parameters we chose $L=2.5, n = 250, m = 200$. The parameters $C_0, \sigma, \kappa, \xi_\pm$ and $\Delta t$ correspond to the Leland numbers $\mbox{Le}=0.85935$ and $\mbox{\underline{Le}}=0.21484$. In Figure~\ref{fig:nonlinearTC-SolutionDelta} we plot the solution $V_{vtc}(S,t)$ and the option price delta factor $\Delta(S,t)=\partial_S V (S,t)$, for $t=0$. The upper dashed line corresponds to the solution of the linear Black--Scholes equation with the higher volatility $\hat \sigma^2_{max}=\sigma^2\left(1-\underline{C}_0\sqrt{\frac{2}{\pi}}\frac{1}{\sigma\sqrt{\Delta t}}\right)$, where $\underline{C}_0=C_0 - \kappa (\xi_+ - \xi_-)>0$, whereas the lower dashed line corresponds to the solution with a lower volatility $\hat \sigma^2_{min}=\sigma^2\left(1-{C_0}\sqrt{\frac{2}{\pi}}\frac{1}{\sigma\sqrt{\Delta t}}\right)$.

\section*{Acknowledgements}

\noindent This research was supported by the European Union in the FP7-PEOPLE-2012-ITN project STRIKE - Novel Methods in Computational Finance (304617).


%
%
%

%
%

\end{document}